\documentclass{aa}  

\usepackage{graphicx}
\usepackage{natbib}
\usepackage[utf8]{inputenc}

\usepackage{txfonts}
\defcitealias{Coccato22}{C22}

\usepackage{xcolor}
\usepackage{orcidlink}

\usepackage{hyperref}
\hypersetup{
    colorlinks=true,
    citecolor=blue,
    linkcolor=blue,
    filecolor=blue,      
    urlcolor=blue,
    }

\newcommand{\msun}{\mathrm{M}_\odot}

\begin{document} 

   \title{Characterizing the Formation and Evolution of S0-galaxies (CaFES-0): Their formation pathways around Galaxy Clusters}

 
 \author{Diego Pallero\orcidlink{0000-0002-1577-7475}
          \inst{1,2}\fnmsep\thanks{\email{diego.pallero@usm.cl}}\and
          Yara~L. Jaff\'e\orcidlink{0000-0003-2150-1130}\inst{1,2}\and
          Amelia Fraser-McKelvie\orcidlink{0000-0001-9557-5648}\inst{3}\and
          Lodovico Coccato\orcidlink{0000-0001-7817-6995}\inst{3}\and
          Facundo A. Gómez\orcidlink{0000-0003-4232-8584}\inst{4}\and
          Yannick Bah\'e\orcidlink{0000-0002-3196-5126}\inst{5,6}\and
          Evelyn J. Johnston\orcidlink{0000-0002-2368-6469}\inst{7}\and
          Ciria Lima-Dias\orcidlink{0009-0006-0373-8168}\inst{4}\and
          Arianna Cortesi\orcidlink{0000-0002-0620-136X}\inst{8,9}\and
          Maria Emilia De Rossi\orcidlink{0000-0002-4575-6886}\inst{10,11}}

   \institute{   Departamento de Física, Universidad Técnica Federico Santa María, Avenida España 1680, Valparaíso, Chile \and
     Millennium Nucleus for Galaxies (MINGAL)\and
     European Southern Observatory, Karl-Schwarzschild-Straße 2,
    Garching, 85748, Germany \and
    Departamento de Astronom\'{i}a, Universidad de La Serena, Av. Juan Cisternas 1200 Norte, La Serena, Chile \and
    School of Physics and Astronomy, University of Nottingham, University Park, Nottingham NG7 2RD, UK \and
    Laboratory of Astrophysics, Ecole Polytechnique Fédérale de Lausanne (EPFL), Observatoire de Sauverny, 1290 Versoix, Switzerland \and
    Instituto de Estudios Astrof\'isicos, Facultad de Ingenier\'ia y Ciencias, Universidad Diego Portales, Av. Ej\'ercito Libertador 441, Santiago, Chile \and
    Instituto de Física, Universidade Federal do Rio de Janeiro, 21941-972, Rio de Janeiro, RJ, Brazil\and
    Universidade Federal do Rio de Janeiro, Observatório do Valongo, Ladeira do Pedro Antônio, 43, Saúde CEP 20080-090 Rio de Janeiro, RJ, Brazil\and
    Universidad de Buenos Aires, Facultad de Ciencias Exactas y Naturales y Ciclo Básico Común, Buenos Aires, Argentina\and
    CONICET-Universidad de Buenos Aires, Instituto de Astronomía y Física del Espacio (IAFE), Buenos Aires, Argentina
             }

   \date{Received September 15, 1996; accepted March 16, 1997}

 
  \abstract
   {The formation pathways of lenticular galaxies (S0s), which lie morphologically between elliptical and spiral galaxies, remain a topic of active research. Environmental effects, merging histories, and pre-processing mechanisms are often proposed as key factors influencing their transformation. However, the relative importance of these processes remains unclear, particularly when compared with other galaxy types.}
   {We investigate and quantify the role of the environment on the formation of S0s.}
   {We use the Hydrangea cosmological zoom-in simulation suite to analyse the environmental histories of S0 galaxies, defined here as central and satellite quenched disk galaxies.}
   {We find that the vast majority (>85\%) of our sample of S0s are satellites in massive haloes (log$_{10}$M$_{200}/$M$_\odot$ > 13), while only $\sim10\%$ are centrals in low-mass haloes. Satellite S0s exhibit a highly quiescent merging history, with $\sim60\%$ experiencing no significant mergers since $z=2$. Centrals show more varied merging histories, although our results may be affected by limited sample size. Contrary to expectations, no clear trends in merger ratios with morphology are observed. However, mergers involving lenticular and spiral galaxies tend to occur in low-density environments and are likely gas-rich, enabling disk reformation. Pre-processing effects in groups are critical, influencing both quenching and morphological transformation.}
   {Our results strongly suggest that S0 galaxies predominantly form from faded/stripped spirals in clusters, with a minority forming via mergers in smaller haloes. These results are in agreement with previous observations of lenticular galaxies around galaxy clusters. }

   \keywords{galaxies:formation -- galaxies:evolution -- galaxies:kinematics and dynamics -- galaxies:interactions -- galaxies:lenticulars -- methods:numerical}
\titlerunning{The formation pathways of S0s}
\authorrunning{D. Pallero et al.}
   \maketitle
%

\section{Introduction}

Galaxies can take many shapes and forms. Since the beginning of the 20$^{th}$ century, astronomers have organised galaxies by their shape and properties, the most famous classification scheme being the Hubble Tuning fork \citep{Hubble26}. In this sequence, we can identify three broad groups of galaxies: spiral (S), elliptical (E), and lenticular (S0) galaxies.
Moreover, we now know that a galaxy's morphology is correlated with some of its more fundamental properties. For example, elliptical/early-type galaxies have little to no star formation activity, older stellar ages, and redder colours \citep[]{Gomez03, Kauffmann04, Poggianti06, Moran07, Blanton09}. On the other hand, spiral/late-type galaxies show higher star formation rates, a population dominated by young stars, and bluer colours \citep[]{Postman05, Fasano15, Haines15, Brough17, Cava17, Liu2019ApJL}.
Lenticular galaxies, also known as S0s, have long fascinated astronomers due to their distinctive structural properties. Characterised by a prominent central bulge and a surrounding disk without a spiral arm structure, lenticular galaxies present a valuable case for studying the transition from actively star-forming galaxies to more quiescent systems.
Although we can count several S0s in the field, they are more commonly found in groups and clusters of galaxies \citep{Dressler80}. 
Moreover, as we approach high-density environments, the number of spiral galaxies decreases, while the number of lenticular galaxies increases. One might therefore suspect that S0 galaxies could be spiral galaxies, whose arms faded due to environmental effects.

Current studies from extensive galaxy surveys have shown that S0s can exhibit a range of photometric, spectroscopic, and kinematic properties that cannot be explained as faded spirals \citep{Tous2025MNRAS}. 
Because of this, understanding the formation and evolution of lenticular galaxies is crucial to unravel the broader context of galaxy morphology and the processes that drive the cessation of star formation in the Universe \citep[e.g.][]{Jaffe11, Johnston14, Johnston21, Fraser-McKelvie18,  Fraser-McKelvie21, Coccato22,  Deeley20, Deeley21}.

Recent advances in observations and simulations have provided new insights into the physical mechanisms that may lead to the development of lenticular morphology, particularly in galaxies that have retained their disks but do not exhibit spiral arms \citep{Deeley21}. The current preferred paradigm is therefore that there are two scenarios for S0 formation: they can form through fading of spirals or via mergers \citep{Emsellem11, Deeley20, Coccato20, Coccato22}.
In the faded spiral scenario, the gas within spiral arms is thought to be removed by environmental effects, such as ram-pressure stripping or strangulation \citep[e.g.,] []{deLucia12, Wetzel13, Wetzel15, Muzzin14, Peng15, Pallero22}. The second formation pathway is through interactions with other galaxies, such as minor and major mergers \citep{Bekki11, Tapia17, Deeley21}. 
Each formation scenario yields distinct properties in the formed S0s.
Those formed from mergers tend to be more massive, older, less metal-rich, and more pressure-supported, with higher S\'ersic \citep{Sersic63} indices and lower ellipticities; they may also exhibit direct evidence of their past mergers, such as shells or stellar streams. 
On the other hand, S0s formed as faded spirals tend to be less massive, with younger stellar populations, be metal-rich, be more rotationally supported, and have lower S\'ersic indices. 
Although these differences can, in principle, be used to infer the formation pathways of observed S0s, this has proven challenging in practice. The main reason for this is that the formation histories of galaxies are more complex than those described in the idealised scenarios above, so that there is no clear threshold in any one property that distinguishes between the two formation pathways unambiguously.

The pilot work of \citet[][C22 hereafter]{Coccato22} addresses this by combining kinematics, morphology, and properties of the stellar populations of 329 S0 galaxies selected from the MaNGA \citep{Manga} and SAMI \citep{SAMI} galaxy surveys, as derived in  \citet{Fraser-McKelvie18, Fraser-McKelvie21} and \citet{Coccato20} to shed light on the predominant formation mechanism of S0 galaxies. Using a clustering algorithm based on properties expected to differ between the two formation mechanisms, they identify their galaxies as either ``faded spiral" or ``merged". Their study found that the ``faded spiral” pathway is the dominant mechanism for forming S0s, and its efficiency increases with the mass of the group or cluster, as well as the local density.
On the other hand, \citet{Deeley21} took a theoretical approach and looked at the different formation mechanisms of S0s at $z=0$ using the magnetohydrodynamical simulations from the IllustrisTNG suite. For this purpose, 500 mock SDSS images of randomly selected IllustrisTNG galaxies were created using the SKIRT radiative transfer code \citep{SKIRT}, adjusted to match the angular size distribution of the SAMI sample. Then, they traced the formation history of the chosen S0s to identify the two main formation mechanisms mentioned above directly. In contrast to \citealt{Coccato22}, only 37$\%$ of the S0s in their sample were formed via gas stripping, i.e., fading of spirals. Moreover, contrary to expectations, they suggest that 57\% of S0 galaxies formed through mergers, indicating that mergers are the primary mechanism of S0 formation.

The discrepant conclusions of \citetalias{Coccato22} and \citet{Deeley21} regarding the dominant formation pathways of lenticular galaxies could be a reflection of the limitations of the datasets used. On the one hand, the sample used by \citet{Deeley21} lacked the massive clusters necessary for a fair comparison with the SAMI data. 
Additionally, the comparison with observational data was limited to the group-size halos from the GAMA survey. 
On the other hand, the work of \citetalias{Coccato22} lacks a sufficiently high number of S0 galaxies to constitute a statistically representative sample, as well as completeness in low-density environments. Moreover, as an observational work, tracing back the assembly histories of the two populations to confirm their origin was not possible.

In this work, we analyse the origin of S0s in the state-of-the-art cosmological Hydrangea zoom-in simulations. This suite is comprised of 24 galaxy clusters, ranging from $10^{14} \leq M_{200}/M_\odot \leq 10^{15.4}$, covering the whole range of galaxy clusters, from low- to very high-mass structures. Additionally, the high-resolution zoom-in region of each simulation extends to $10r_{200}$ from the centre of each cluster and therefore captures a variety of environments beyond the clusters themselves. 
We select S0s using a combination of morpho-kinematic measurements and their specific star formation rate (sSFR) to identify lenticular galaxies according to a physical definition, namely: rotationally-supported galaxies with little to no star formation.

This paper is organised as follows: in Section~\ref{sec:methods}, we summarise the properties of the Hydrangea simulations and the methodology used to measure the morpho-kinematic properties for classifying our galaxies. In Section~\ref {sec:galaxies}, we perform the morphological classification of our different subsamples and construct a control sample to compare our results. In Section~\ref{sec:results}, we study the environmental dependence of different formation pathways and their impact on their merging history. Additionally, we put our results in the context of comparing with a subset of observational data. Our main conclusions are summarised in Section~\ref{sec:summary}. 

The simulation used on this work assumes the same flat $\Lambda$CDM cosmology used in the original \textsc{EAGLE} project \citep{Schaye15, Crain15}, specifically those of \citet{Planck14}, being: $\Omega_\Lambda = 0.693$, $\Omega_m = 0.307$, $\Omega_b = 0.04825$, $\sigma_8 = 0.8288$, $Y = 0.248$, and $H_0 = 67.77$ km s$^{-1}$ Mpc$^{-1}$.

\section{Methods}
\label{sec:methods}
\subsection{The \textsc{hydrangea} Simulations}

The Hydrangea simulations \citep[see also \citealt{Barnes17}]{Bahe17} are a suite of 24 zoom-in simulations of massive galaxy clusters based on the successful galaxy formation model developed for the EAGLE project \citep{Schaye15}. We summarise their most relevant aspects here, but refer the reader to the aforementioned works for comprehensive details on the simulation framework and subgrid physics.

The simulated clusters were selected from an $N$-body simulation with a comoving box side of 3.2 comoving Gpc \citep{Barnes17a}, with halo masses in the range of $14 \leq \log_{10}(M_{200}^{z=0}/\mathrm{M}_\odot) \leq 15.4$ and without an even more massive nearby cluster. Zoom-in initial conditions were constructed to achieve a high-resolution region of at least 10 $r_{200c}$ around each cluster, robustly capturing galaxies evolving across the wide range of local environments on the far outskirts of clusters. The high-resolution region uses particle masses of $m_{\mathrm{DM}} = 9.7 \times 10^6 \, \mathrm{M}_\odot$ for dark matter and $m{\mathrm{gas}} = 1.8 \times 10^6 \, \mathrm{M}_\odot$ for baryons (gas and stars). The gravitational softening length was fixed at $\epsilon = 2.66$ ckpc for redshifts $z \geq 2.8$, and $\epsilon = 0.7$ physical kpc for $z < 2.8$.

The resimulations were then performed with an updated version of the Tree-PM Smoothed Particle Hydrodynamics (SPH) code \textsc{GADGET-3} \citep{Springel05}, using the `\textsc{Anarchy}' hydrodynamics scheme \citep{Schaye15, Schaller15} and the `AGNdT9' variant of the \textsc{EAGLE} galaxy formation model as presented in \citet{Schaye15}. As described in detail by \citet{Schaye15}, the subgrid physics prescriptions of the EAGLE model include radiative cooling and photo-heating following \citet{Wiersma09}, stochastic star formation with a metallicity-dependent threshold \citep{Schaye04, Schaye08}, a pressure floor corresponding to $P \propto \rho^{4/3}$ imposed on gas with $n_\mathrm{H} \geq 10^{-1}\,\mathrm{cm}^{-3}$ to prevent the formation of an inadequately modeled cold gas phase, mass and metal enrichment from stellar evolution \citep{Wiersma09b} and energy feedback from star formation \citep{Vecchia12} and accreting supermassive black holes \citep{Rosas-Guevara15, Schaye15}.

The galaxy formation model has been extensively validated in the \textsc{EAGLE} project. \citet{Schaye15} demonstrated that it reproduces a broad range of galaxy properties not explicitly calibrated for, in addition to other properties such as neutral gas fractions \citep{Bahe16, Crain17}, a key parameter for modeling star formation. \citet{Barnes17} further demonstrated that, although the clusters within the simulation are more gas-rich than observed, they can capture some intra-cluster medium (ICM) properties, such as the gas fraction–mass relation and metallicity, within reasonable agreement. The X-ray and Sunyaev-Zel'dovich (SZ) observables (e.g., temperature, luminosity, and mass relations) also match observational constraints.

Simulation outputs are stored at 30 snapshots over the redshift range from $z = 14$ to $z = 0$. In each of them, gravitationally bound structures are identified with a two-step post-processing procedure. 
First, dark matter halos are found using a friends-of-friends (FoF) algorithm, with a linking length set to $b = 0.2$ times the mean inter-particle separation. Baryonic particles (gas and stars) are then associated with the FoF group of their nearest dark matter particle. FoF groups containing fewer than 32 dark matter particles are excluded from further analysis due to insufficient resolution. In a second step, gravitationally bound substructures (subhaloes) within each FoF halo are identified by the \textsc{SUBFIND} code \citep[see also \citealt{Springel01}]{Dolag09}. The algorithm identifies local overdensities and applies a binding-energy criterion to distinguish self-bound systems, accounting for both dark matter and baryonic components.

Based on the \textsc{Subfind} catalogues in individual snapshots, galaxy tracks across cosmic time were then built with the \textsc{SPIDERWEB} algorithm \citep[see Appendix A in][]{Bahe19}. This code has been designed to robustly track the evolution of galaxies in dense environments by identifying the best descendant subhalo for each galaxy, even when subhaloes exchange a substantial fraction of their particles.
This enables us to reliably trace the evolution and merger history of galaxies, even in the complex dynamical conditions of galaxy clusters.

\subsection{IllustrisTNG}
Additionally, to put our results into context, we will revise our main results in the IllustrisTNG cosmological volume. The IllustrisTNG suite \citep[TNG;][]{Marinacci18, Naiman18, Nelson18, Pillepich18, Springel18}, is a set of magneto-hydrodynamical cosmological simulations with a comprehensive galaxy formation model including star formation, feedback, black-hole physics, and metal enrichment \citep[see][for details]{Pillepich18a}. We focus on TNG100, which provides an optimal compromise between volume and resolution, enabling a representative galaxy sample across diverse environments while resolving both kinematics and morphology. TNG100 has baryonic and dark matter particle masses of $m_{\rm bar}\sim1.4\times10^{6}{\rm M_\odot}$ and $m_{\rm DM}\sim7.5\times10^{6}{\rm M_\odot}$, respectively, being comparable with the resolution level of Hydrangea.
Simulations were run with the \textsc{arepo} code \citep{Springel10}, adopting a flat $\Lambda$CDM cosmology consistent with Planck results \citep{Planck16}. TNG100 reproduces key observables, including galaxy colour bimodality, the mass-metallicity relation, galaxy sizes, and intra-cluster medium metal distributions \citep[e.g.,][]{Nelson18, Torrey18, Genel18, Vogelsberger18, Pillepich18a}.
The public catalogues provide galaxy fluxes in multiple bands \citep{Nelson18}. Each star particle is treated as a simple stellar population, with spectral properties modelled using FSPS, Padova isochrones, the MILES library, and a Chabrier IMF \citep{Chabrier03}. Galaxy magnitudes are calculated by summing the magnitudes of all star particles, accounting for dust and nebular emission corrections.
Overall, TNG100 provides an ideal laboratory for studying galaxy evolution and morphological transformation across diverse environments, though it lacks the massive clusters found in Hydrangea. For this reason, we develop the bulk of our results in the Hydrangea simulation throughout the article. Nonetheless, we include a similar analysis developed for TNG in Appendix \ref{sec:TNG}. 

\subsection{Galaxy selection}

For our study, we select only well-resolved galaxies with stellar mass $M_{\star} > 10^{10}\,\mathrm{M}_\odot$; these are composed of at least several thousand star particles and therefore allow us to determine their morphology robustly. 

To exclude contamination from low-resolution particles outside the high-resolution region, we only consider galaxies that are within $10\,r_{200}$ of the central cluster in each simulation and are at least 5 Mpc from the nearest low-resolution `boundary' particle at $z = 0$. In total, our sample includes 1592 galaxies per morphological type. 

By selecting galaxies from a zoom-in simulation, our sample includes many galaxies in cosmologically rare environments, such as massive galaxy clusters. Extending our selection to $10\,r_{200}$ from the cluster centre means that we also include many non-cluster galaxies, so that we can study the formation of S0s over a vast gamut of local environments. Nevertheless, it is essential to keep in mind that our sample is biased towards large-scale overdensities and hence not representative of the full S0 population.

\subsection{Kinematic measurements}
To identify simulated S0 galaxies, we need to distinguish between disks and spheroids. We use the kinematic properties of galaxies for this purpose, following the framework proposed by
\citet{Lagos17} that includes measurements of the stellar velocity dispersion ($\sigma$), specific angular momentum ($j_\star$), rotational velocity ($V_{\rm rot}$), and the stellar angular momentum ($\lambda_R$).
Unless stated otherwise, all kinematic and structural properties are computed within an aperture of three stellar half-mass radii, which we adopt to select particles belonging to the bulge and disks of galaxies. This aperture ensures a consistent measurement region across galaxies, accounting for each galaxy's mass and size while focusing on the central parts where the signal is most reliable \citep[][]{Naab14, Lagos17, Pallero25}.

We begin by calculating the specific angular momentum of the stellar component, using
\begin{equation}
    j_{\star} = \frac{\sum_i m_i (r_i - r_{\rm{COM}})\times (v_i - v_{\rm{COM}})}{\sum m_i}.
\end{equation}
Here, ${r}_i$ and ${v}_i$ are the position and velocity of the $i^\mathrm{th}$ stellar particle, and $\mathbf{r}_{\rm COM}$ and $\mathbf{v}_{\rm COM}$ are the position and velocity of the galaxy’s centre of mass, respectively. The latter are measured from the positions of all stellar particles within 3$r_{1/2}$ from the position of the most bound particle in the subhalo, where $r_{1/2}$ is the radius enclosing half of the total stellar mass of the galaxy.
From $j_\star$, we derive the rotational velocity at radius $r$ as
\begin{equation}
    V_{\rm rot}(r) \equiv \frac{\left| j_\star(r)\right|}{r}. 
    \label{eq:vrot}
\end{equation}

Additionally, we measure the velocity dispersion perpendicular to the mid-plane of the disk by projecting stellar velocities along the direction of the total angular momentum vector $\mathbf{L}_\star$, computed using all gravitationally bound stellar particles. 
The 1D velocity dispersion perpendicular to the disk plane is given by
\begin{equation}
    \sigma_{1,D} = \sqrt{\frac{\sum_i m_i (\Delta v_i cos\theta_i)2 }{m_i}},
    \label{eq:sig1d}
\end{equation}
where $\Delta v_i = |\mathbf{v}i - \mathbf{v}_{\rm COM}|$ is the velocity relative to the centre of mass, and $\cos\theta_i = {({v}_i - {v}_{\rm COM}) \cdot {L}_\star}/({|{v}_i - {v}_{\rm COM}||{L}_\star|)}$ projects the velocity along the angular momentum vector.

From these quantities, we compute the $\lambda_R$ parameter introduced by \citet{Emsellem07} to quantify the degree of rotational support in galaxies. This parameter was initially introduced as a lightweight measure of the rotational support of galaxies, but previous simulation work has shown that similar results are obtained with its mass-weighted analogue, as used here \citep[][]{Naab14, Lagos17, Pallero25}.

Following \citet{Naab14}, we compute $\lambda_R$ by summing over concentric radial bins,
\begin{equation}
    \lambda_R = \frac{\sum_{i=1}^{N(r)} m_{\star,i}r_i V_{rot}(r_i)}{\sum_{i=1}^{N(r)}\sqrt{V_{rot}^{2}(r_i) + \sigma_{1D,\star}^{2}(r_i)}},
\end{equation}
where $N(r)$ is the total number of radial bins within the aperture, $m_{\star,i}$ is the stellar mass enclosed in each bin. $V_\mathrm{rot} (r_i)$ and $\sigma_\mathrm{1D} (r_i)$ are calculated from equations \eqref{eq:vrot} and \eqref{eq:sig1d}, respectively based on the particles within each bin. 

While $\lambda_R$ is inherently sensitive to binning resolution, using a fixed aperture of 3 stellar half-mass radii across the sample provides consistency in the analysis. Depending on the galaxy's stellar mass, this value ranges between 3 and 30 kpc.

\subsection{Galaxy ellipticities}
In addition to kinematics, we estimate the stellar ellipticity of each galaxy within the same aperture from the moment-of-inertia tensor of the stellar component. This is defined as
\begin{eqnarray}
    I(r)_{xx} &=& \sum_{i} m_{i}(y^{2} + z^{2}) \nonumber \\
    I(r)_{xy} &=& -\sum_ix_iy_i,
\end{eqnarray}
with similar expressions for the other components due to the tensor’s symmetry. After computing the full inertia tensor, we diagonalise it and sort the eigenvalues ($\lambda_1 \leq \lambda_2 \leq \lambda_3$) to determine the principal axes of the galaxy. These axes are then used to compute the semi-axes lengths as shown in \citet{Candlish18}:
\begin{equation}
\begin{split}
    a &= \sqrt{\lambda_2+\lambda_3-\lambda_1} \\
    b &= \sqrt{\lambda_1+\lambda_3-\lambda_2} \\
    c &= \sqrt{\lambda_1+\lambda_2-\lambda_3}.
\end{split}
\end{equation}
From the major and minor axes $a$ and $c$ obtained in this way, we calculate the ellipticity of the galaxy as
\begin{equation}
    \varepsilon = 1 - \frac{c}{a}.
\end{equation}


\section{Galaxy Classification}
\label{sec:galaxies}
There is no unambiguous quantitative definition of lenticular galaxies in the literature. Different authors have identified them in various ways, initially purely visually (e.g., \citealt{Hubble26, deVaucouleurs59}). Still, the advent of IFU instruments with large fields of view \citep[e.g.,][]{Emsellem07, Emsellem11, Capellari11, SAMI} brought more sophisticated identification techniques that combine morphology and kinematics.
In recent years, automated methods including machine learning and deep learning algorithms have also been used to categorise galaxies by morphological type \citep[e.g.,][]{DLpaperTNG, Walmsley23, Kolesnikov24, Sampaio25}. In light of this situation, we use a simple theoretical definition of a lenticular galaxy: a disk-dominated galaxy with little or no star formation.

\subsection{Morphological classification}
To define a galaxy as disk, we use the threshold proposed by \citet[][]{Emsellem07}, $\lambda/\sqrt{\varepsilon} > 0.31$, to separate galaxies that are rotationally supported from spheroids. This threshold has been used in hydrodynamic simulations by several previous studies to confirm its applicability
\citep[][]{Naab14, Lagos17, Pallero25}. 

\begin{figure}
\centering

\includegraphics[width=0.5\textwidth]{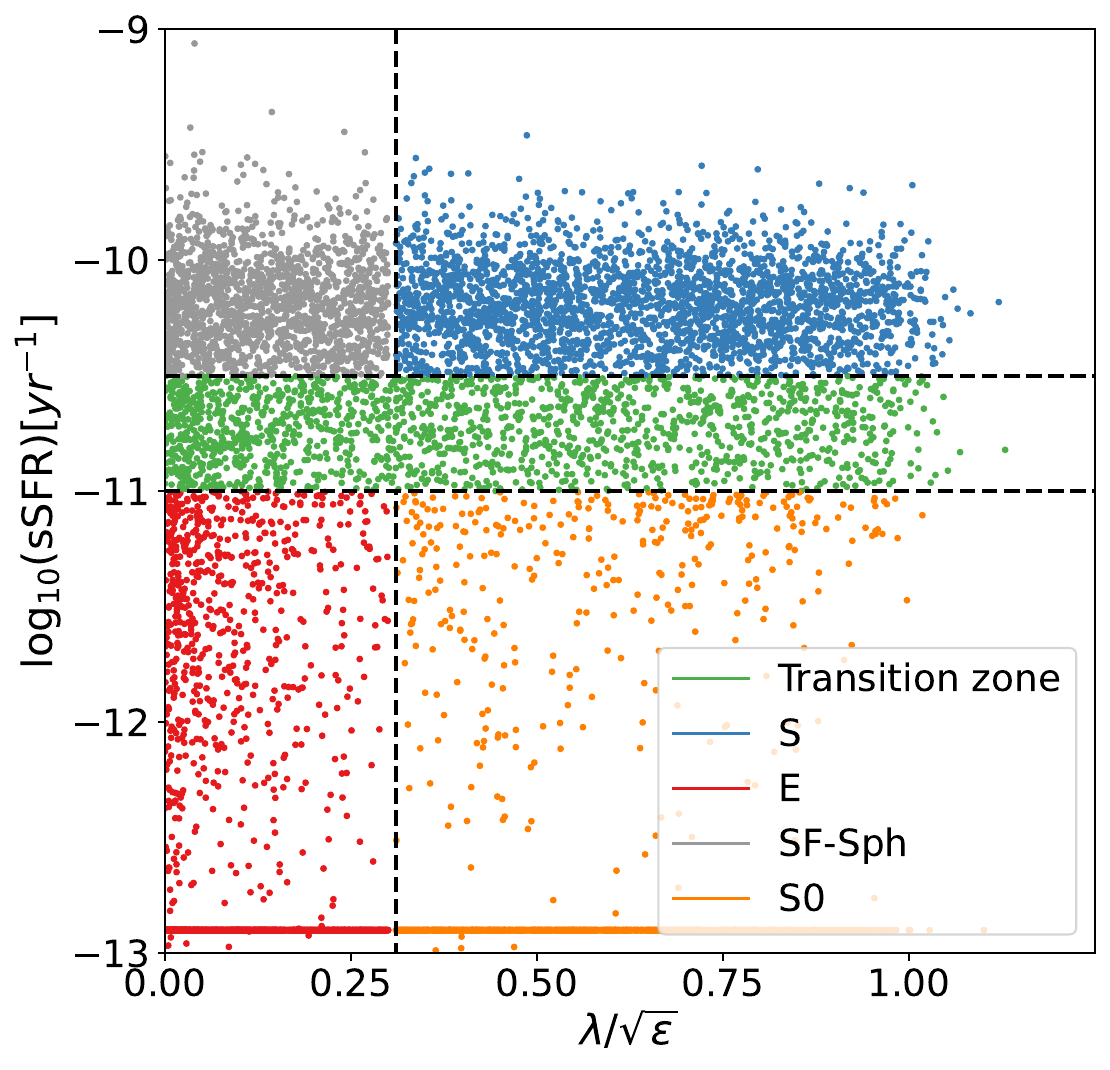}

\caption{Specific star formation rate, as a function of the kinematical properties of the galaxies in our sample.
The dashed lines divide galaxies into five regions based on their kinematical support and sSFR. Galaxies are divided as spirals (S: blue dots, star-forming disks), lenticulars (S0: orange, non-star-forming disks), star-forming ellipticals (SF-Sph: gray, star-forming spheroids), ellipticals (E: red dots, non-star-forming spheroids), and a transition zone (green dots, -11 < log$_{10}$sSFR[$yr^{-1}$] < -10.5).}

\label{fig:sel_all}
\end{figure}

To exclude galaxies with significant ongoing star formation, we impose a threshold in specific star formation rate (sSFR) of $< 10^{-11}\,\mathrm{yr}^{-1}$, the same as used by many previous $z \sim 0$ analyses \citep[e.g.][]{Muzzin12, Wetzel12, Dave19, Pallero19, Pallero22}. This definition of `quenched' corresponds approximately to a galaxy that cannot double its stellar mass in a Hubble time. To separate quenched and star-forming galaxies more clearly, we define the latter as having sSFR $> 10^{-10.5}\,\mathrm{yr}^{-1}$.
It should be noted that, as discussed in \citet[][]{Pallero22}, the sSFR threshold used to define galaxies as passive (within $\pm 0.25$ dex) has little impact on the overall results on how galaxies in this mass range reach their quenching state.
In total, these criteria separate our galaxy sample into five populations, as illustrated in Fig.~\ref{fig:sel_all}. These include star-forming disks (`spirals', blue, $N = 2737$) with sSFR $>10^{-10.5}\,\mathrm{yr}^{-1}$ and $\lambda_R/\sqrt{\varepsilon} > 0.31$, quenched disks (`S0s', orange, $N = 1727$) with sSFR $<10^{-11}\,\mathrm{yr}^{-1}$ and $\lambda_R/\sqrt{\varepsilon} > 0.31$, and quenched spheroids (`Ellipticals', red, $N = 2238$) with sSFR $<10^{-11}\,\mathrm{yr}^{-1}$ and $\lambda_R/\sqrt{\varepsilon} < 0.31$. In addition, there are star-forming spheroids (grey, top-left) and a `transition zone' between the star-forming and quenched populations, but we do not consider these further here.

\subsection{The need for a control sample}

\begin{figure}
\centering
\includegraphics[width=0.5\textwidth]{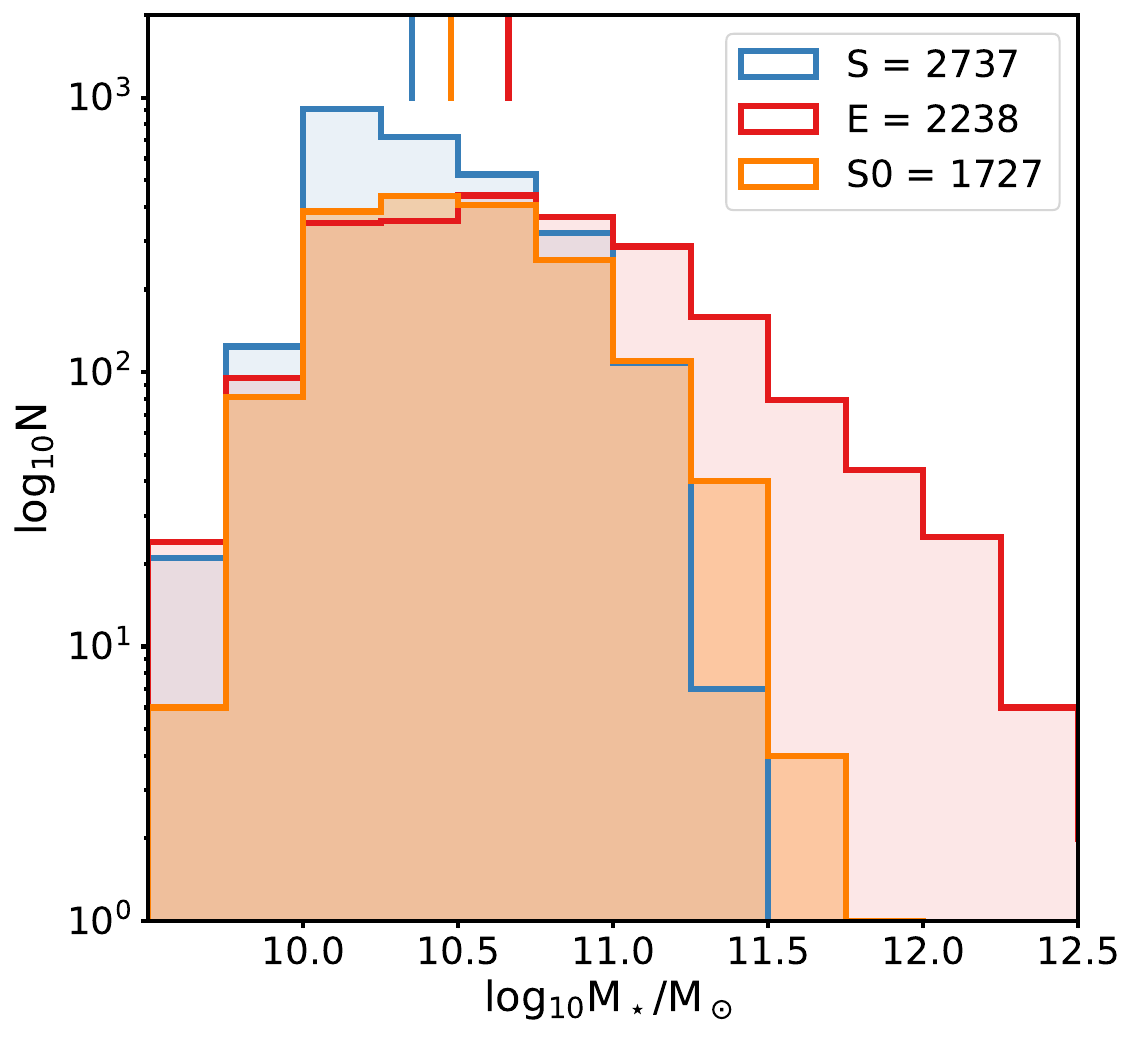}

\caption{Stellar mass distribution of galaxies split by morphological type as spirals (blue), lenticular (orange), and elliptical (red) galaxies.
As shown in this Figure, different morphologies exhibit distinct stellar mass distributions. For this reason, we decided to use a stellar-mass control sample to conduct our study.
}

\label{fig:Mstar_dist}
\end{figure}

To understand the mechanisms that lead to the formation of S0 galaxies, we must compare their properties and formation histories with those of a control sample of S and E galaxies. A direct comparison of all three samples is, however, not useful because they have different stellar mass distributions (see Fig.~\ref{fig:Mstar_dist}), so that one would run the risk of confusing trends with mass and morphology. In particular, spirals are biased to slightly lower mass than S0s---the medians are offset by $\approx$0.1 dex, and there is a noticeable excess of spirals at $10^{10} \lesssim M_\star\,/\,\mathrm{M}_\odot \lesssim 10^{11}$---while ellipticals dominate the high-mass end. These trends are consistent with previous work based on representative simulation volumes (e.g.~\citealt{Clauwens18}).

To make an unbiased comparison between S0, E, and S galaxies, we therefore create three mass-matched subsamples. First, we discard the most massive galaxies ($M_\star > 10^{11.25}\,\mathrm{M}_\odot$) since they are dominated by ellipticals and any S0s that do lie in this range are likely atypical. For each of the 1525 remaining S0s, we search for S and E galaxies that closely match its stellar mass (within 0.01 dex) and then choose one random unique analogue of each type (S, E) from this matched subsample.
If there are no analogues that fulfil the requirements above in each morphological sample (S, E), the galaxy is discarded. The final matched subsamples we use for our analysis below, therefore, contain 1446 galaxies, losing only 16\% of the initial S0 sample.

\section{Results}
\label{sec:results}

\subsection{Environment of S0s}
\begin{figure}
\centering
\includegraphics[width=0.5\textwidth]{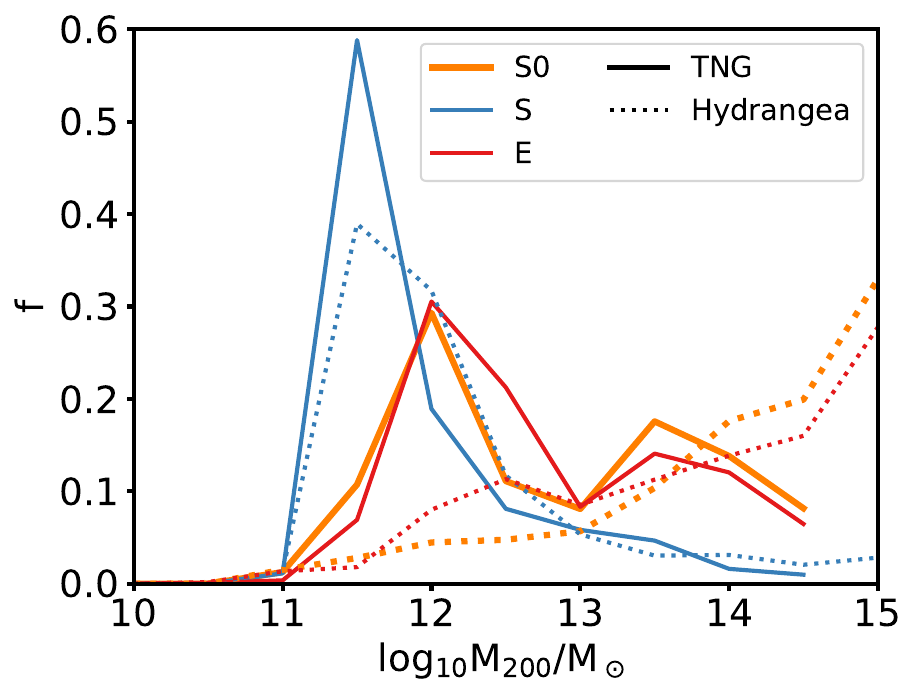}

\caption{Distribution of Spirals (blue), Lenticular (orange), and Elliptical (red) galaxies as a function of the halo mass  M$_{200}$ in which they reside at z=0, for Hydrangea (dashed) and TNG100 (solid). In the Hydrangea simulation, spiral galaxies reside preferentially in low-density environments, contrary to early-type galaxies (lenticular and elliptical) that show a strong preference to reside within dense environments ($log_{10}M_{200}/M_\odot \geq 13$). Nonetheless, there is a non-negligible population of early-type galaxies ($\sim 20\%$) inhabiting haloes with ($12 \leq log_{10}M_{200}/M_\odot \leq 13$). The overall fractions change when considering a cosmological volume of the size of TNG100, with more early-type galaxies residing in low- and intermediate-mass haloes.}

\label{fig:M200_dist}
\end{figure}

As the local environment of a galaxy is known to play a key role in its evolution (e.g.~\citealt{Dressler80, Peng10, Bahe13, Pallero19, Pallero22}), we begin by comparing the distribution of host halo masses for each of the three morphological types. This is shown in Fig.~\ref{fig:M200_dist} and clearly reveals a major dichotomy between spirals (blue), which mostly occupy low-mass haloes ($M_\mathrm{200c} \lesssim 10^{13}\,\msun$), and S0/E galaxies that follow very similar distributions and are most commonly found in clusters $M_\mathrm{200c} > 10^{14}\,\msun$). 
Nevertheless, we note that the host mass distributions of S0 (and also E) galaxies are broad, and 13.5\% of S0s (and 22.6\% of E) reside in haloes typical of low- and intermediate-density environments ($M_\mathrm{200c} < 10^{13}\,\msun$). 
Additionally, this Figure compares what is expected for a cosmological simulation, such as TNG100 (solid line), with our sample of galaxies from the Hydrangea simulation (dashed line). As stressed before, the Hydrangea results are heavily biased toward massive haloes by construction, oversampling the extreme environments of massive galaxy clusters. As TNG simulates a cosmological volume, the fraction of S0/E galaxies residing in low-mass haloes is larger compared with Hydrangea, with 52\% of S0s and 59\% of E inhabiting low- and intermediate density environments ($M_\mathrm{200c} < 10^{13}\,\msun$). 
In Appendix \ref{sec:TNG}, there is a more detailed analysis of our results found in TNG. It is worth mentioning that most S0s found in TNG ($\sim 60\%$) experience no significant mergers at $z<2$. 
We left the detailed analysis of the main formation pathways of S0s galaxies in a cosmological volume containing massive clusters for a future study.

\begin{figure}
\centering
\includegraphics[width=0.5\textwidth]{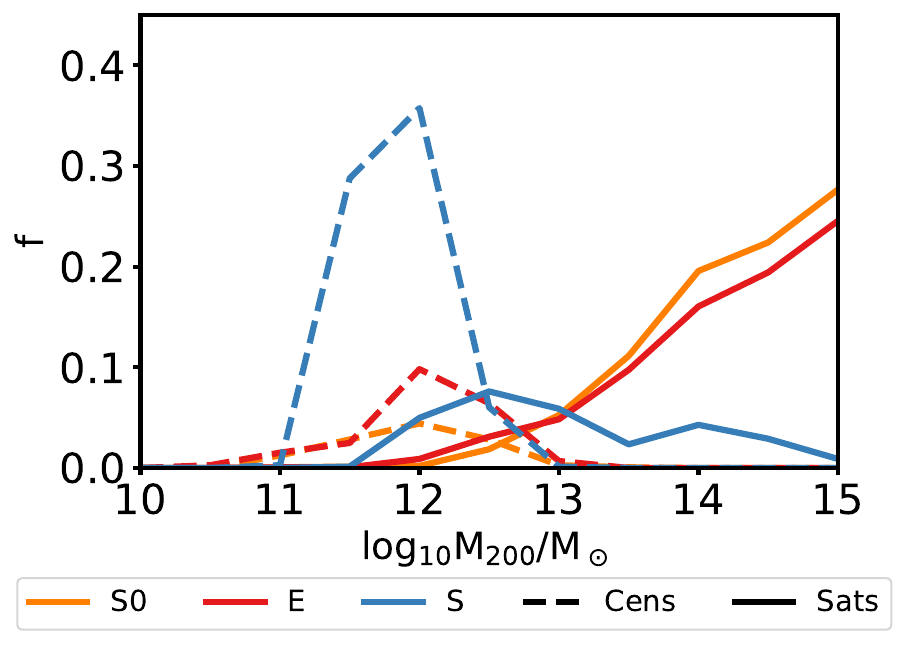}

\caption{Distribution of Spirals (blue), Lenticular (orange), and Elliptical (red) galaxies as a function of the mass of the halo M$_{200}$ in which they reside at z=0. Galaxies are categorised as satellites (solid) or central (dashed). }

\label{fig:sat_cen-split}
\end{figure}

A second key factor known to affect the evolution of a galaxy is whether it is a central---i.e.~the most massive galaxy that sits more or less at rest near the centre of its halo---or a satellite that orbits a more massive central. We distinguish these two sub-populations in Fig.~\ref{fig:sat_cen-split}, where we again show the distribution of S, S0, and E galaxies by halo mass but this time separately for centrals (dashed lines) and satellites (solid lines).

For all three morphological types, satellites occupy systematically more massive haloes; this is not surprising given that the number of satellites per halo increases strongly with halo mass. Nevertheless, there is a clear difference between spirals, where even satellites are mostly restricted to $M_\mathrm{200c} \lesssim 10^{13}\,\msun$, and S0/E galaxies for which the fraction is highest at the high-mass end. Centrals of all three types occupy a broadly similar mass range, $10^{11} \lesssim M_\mathrm{200c}/\msun\,\lesssim 10^{13}$. These findings suggest that satellite-specific processes contribute to the formation of S0s. Still, they cannot be held solely responsible and do not seem to account for the morphological difference between S0 and E galaxies.

\subsection{Merging history}
\begin{figure}
\centering
\includegraphics[width=0.5\textwidth]{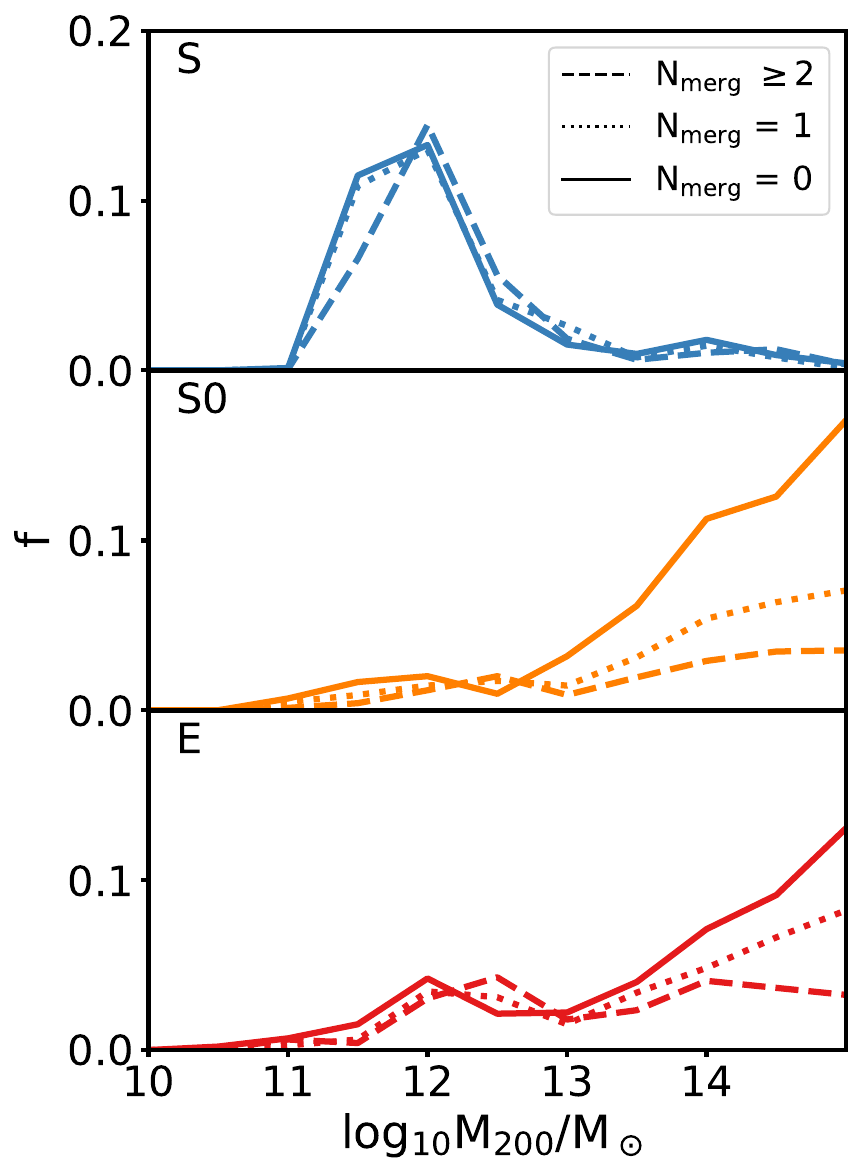}
\caption{Distribution of Spirals (top, blue colours), Lenticular (centre, orange colours), and Elliptical (bottom, red colours) galaxies as a function of the mass of the halo M$_{200}$ in which they reside at z=0. Galaxies are split by the number of significant mergers that they experienced (Merger ratio = $M_{\star, sat} /M_{\star, host}$ > 0.1) at $z<2$, as $N_{merg} \geq 2$ (dashed); $N_{mergers} = 1$ (dotted) and $N_{mergers} = 0$ (solid).}

\label{fig:M200_merg}
\end{figure}

We now turn our attention to the second channel proposed as a pathway to S0 formation: galaxy mergers. In Fig.~\ref{fig:M200_merg}, we compare again the distributions of halo masses for the three morphological samples. Still, this time, split galaxies by the number of significant mergers that they have experienced since $z = 2$. We distinguish between galaxies with N$_{mergers} = 0, 1$ or $\geq2$ in continuous, dotted, and dashed lines, respectively, where significant mergers are defined as the interaction between galaxies if the stellar mass ratio $M_{\star, satellite}/M_{\star, host} > 0.1$.

While the small subset of S0 galaxies in low-mass haloes---which, as we have seen above, are mostly centrals---have a diverse range of merger histories, those in massive clusters have most commonly not undergone any merger in the last 10 Gyr. Only 14\% of S0s in haloes with $M_\mathrm{200c} > 10^{14}\,\msun$ have had two or more mergers, and less than 10\% of the overall sample.

Comparing these trends to those of ellipticals, we observe a similarly low fraction of galaxies with two or more mergers (mostly centrals in low-mass halos). The fraction of galaxies experiencing multiple mergers (dashed lines), however, is notably lower in the most massive haloes, with a higher fraction experiencing a single merger or none. For spirals, we do not find any significant difference in merging history with halo mass, with an approximately even split between the three merging categories from the least to the most massive haloes.

It is worth noting that, when comparing our results with those obtained from TNG (Appendix~\ref{sec:TNG}), the distribution of E/S0 galaxies differs quantitatively from that in Hydrangea, not only in terms of environment, but also in the number of mergers experienced by the galaxies. In particular, E/S0 galaxies with no significant mergers exhibit two prominent peaks at low ($M_{200}/M_\odot \sim 10^{12}$) and intermediate ($M_{200}/M_\odot \sim 10^{13.5}$) halo masses. A similar, although less pronounced, bimodality is also present among galaxies that experienced $\geq,2$ mergers since $z=2$.
As expected, this behaviour is a consequence of the oversampling of massive haloes in Hydrangea. Nonetheless, it should be noted that the qualitative trends remain consistent across simulations; taking this into account, we find that the distribution of satellite E/S0 galaxies increases with $M_{200}$, while galaxies with $N_{\rm merger}=0$ dominate across all environments.
Additionally, we highlight the existence of a non-negligible population of S0 galaxies that did not undergo any significant merger and reside as centrals of low-mass haloes. These systems are likely quenched through mechanisms other than ram-pressure stripping, such as starvation or AGN feedback. A detailed analysis of this population, however, is deferred to future work using a larger cosmological volume.

\begin{figure*}
\centering
\includegraphics[width=0.9\textwidth]{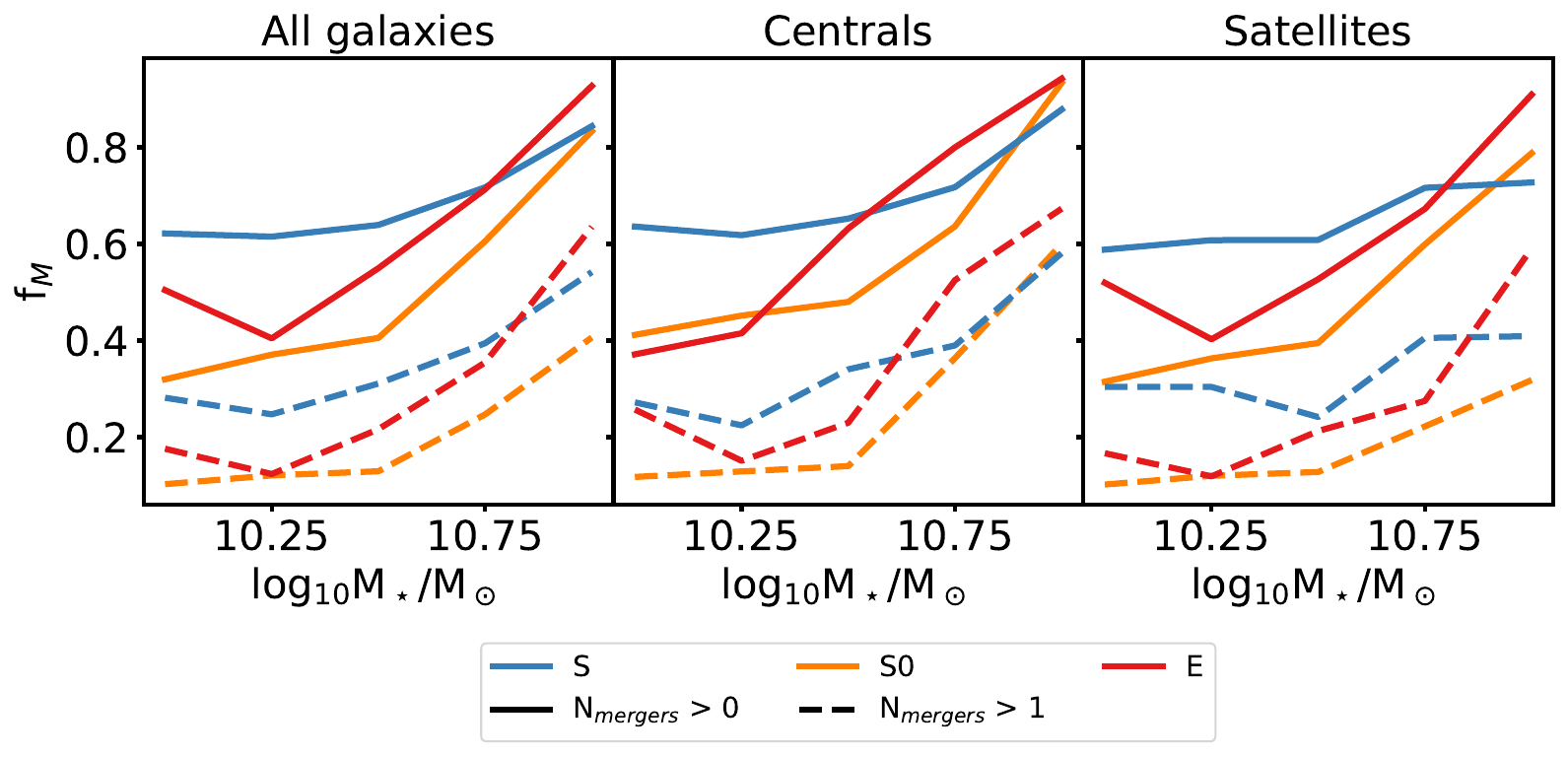}
\caption{Fraction of galaxies that experienced mergers as a function of stellar mass. Blue, red, and orange lines represent spiral, elliptical, and lenticular galaxies, respectively. Solid lines indicate the fraction of galaxies that experienced at least one merger, while dashed lines correspond to those with at least two mergers.
Contrary to expectations, spiral galaxies show a relatively high merger fraction, especially at lower stellar masses. At the high-mass end, elliptical galaxies exhibit the highest incidence of mergers across all morphologies. Interestingly, lenticular galaxies do not appear to require multiple mergers for their formation and consistently show the lowest merger fractions at all stellar mass bins.}

\label{fig:relmerg}
\end{figure*}
For a complementary view of the importance of mergers, we compare in Fig.~\ref{fig:relmerg} the fraction of galaxies that have experienced mergers as a function of stellar mass for the different morphologies, now explicitly distinguishing between centrals and satellites. The primary trend is that more massive galaxies are more likely to have undergone at least one merger; this holds across all categories except for the minority of satellite spirals. To the second order, however, we find that S0s are systematically less affected by mergers at almost all stellar masses, regardless of whether they are central or satellite galaxies.  
This finding is at odds with previous expectations that mergers should play a central role in the formation of S0 galaxies, as shown in \citet{Deeley21}. It is particularly noteworthy that around half of all central S0s with $M_\star \lesssim 10^{10.5}\,\msun$ have not experienced any mergers and can therefore have been formed neither through mergers nor via environmental effects.

In contrast, elliptical and, especially, spiral galaxies exhibit a higher frequency of impactful mergers. These results suggest that mergers have been more influential in shaping the present-day spiral population. 
While this may be surprising, we have found that the mergers experienced by spiral galaxies tend to be minor and gas-rich. This kind of merger enables spirals to either rebuild or maintain their disk morphology following such interactions, as shown in \citet{Pallero25}. 

\begin{figure*}
\centering
\includegraphics[width=0.9\textwidth]{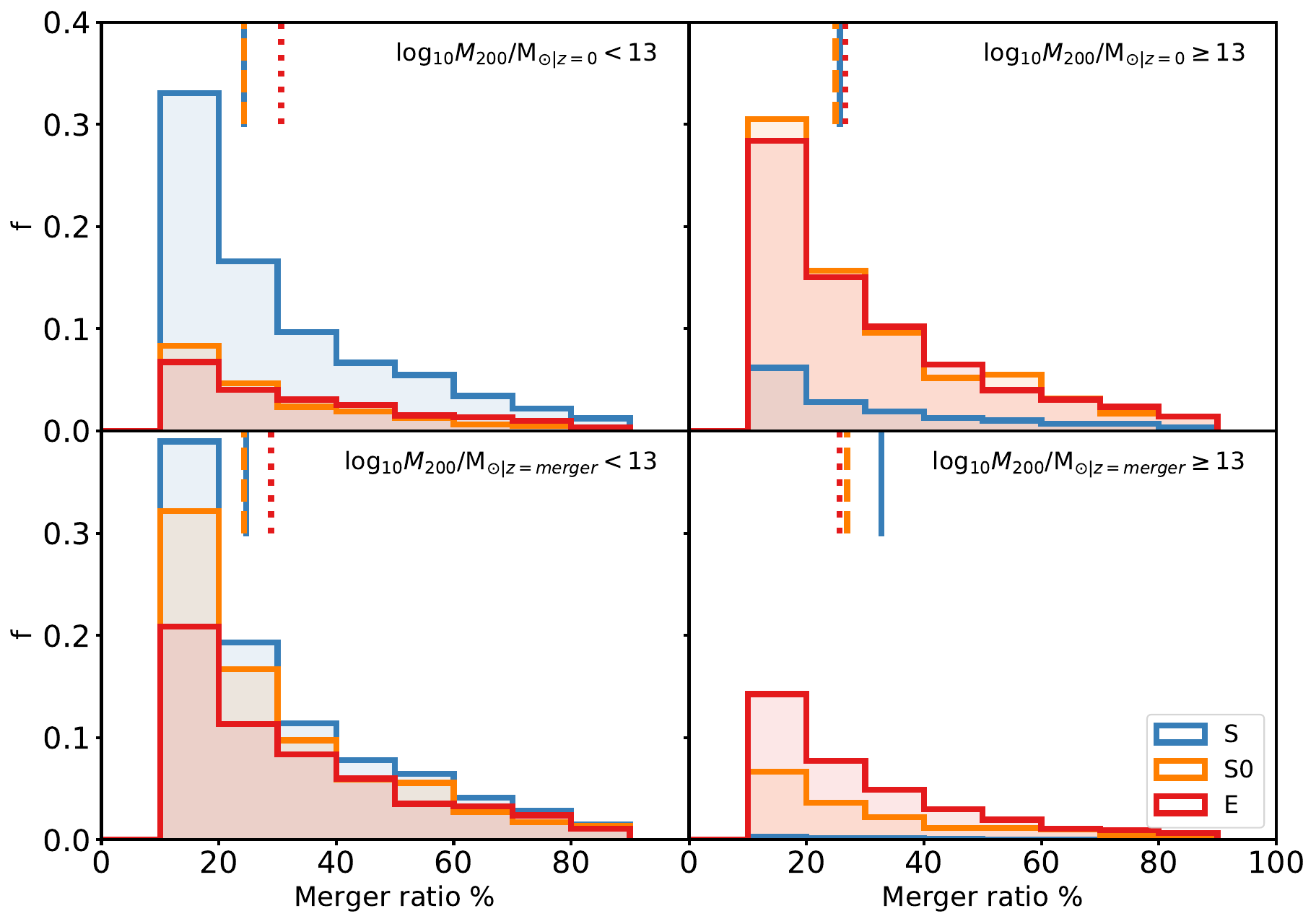}

\caption{Merger ratio distribution for galaxies residing in massive (log$_{10}M_{200}/$M$_\odot \geq 13$; right panels) and low-mass (log$_{10}M_{200}/$M$_\odot < 13$; left panels) environments. The top and bottom panels split the sample by the environment in which galaxies reside at $z=0$ (top panels) and at $z=z_{merger}$ (bottom panels). Spiral (blue), lenticular (orange), and elliptical (red) galaxies are shown, with the median of each distribution indicated in the upper part of each panel by solid, dashed, and dotted lines, respectively.}
\label{fig:merger_ratio_zmerg}
\end{figure*}

For galaxies that have experienced at least one significant merger at $z < 2$, we analyse the distribution of merger ratios in Fig.~\ref{fig:merger_ratio_zmerg}. Each panel displays the fraction of mergers as a function of merger ratio, defined as the stellar mass ratio between the two merging galaxies, before the start of the merging process, with different colours representing different morphological types (orange, red, and blue for S0s, E, and S galaxies, respectively). The two left-hand panels show galaxies in low-mass haloes (log$_{10}$M/M$_\odot < 13$), while their counterparts in more massive haloes are shown in the right-hand panels.
The top row classifies galaxies based on their environment at $z=0$, while the bottom row splits by their environment and the time of each merger ($z_\mathrm{merger}$).

This Figure reveals that mergers in spiral galaxies are predominantly minor, with spiral galaxies experiencing the largest number of minor mergers, followed by lenticular galaxies. This trend is consistent across environments and suggests minor mergers play a role in disk evolution, particularly for late-type galaxies.
A key result emerges when considering the environmental context of S0 mergers. At 
$z=0$, the environmental distribution of lenticular galaxies resembles that of elliptical galaxies, with many residing in high-density environments such as groups or clusters. Interestingly, however, when we examine the environment at the time of merger, lenticular galaxies more closely follow the distribution of spiral galaxies; that is, most of their mergers occurred in low-density environments, typically outside of galaxy clusters, and were predominantly minor.

\begin{figure*}
\centering
\includegraphics[width=0.9\textwidth]{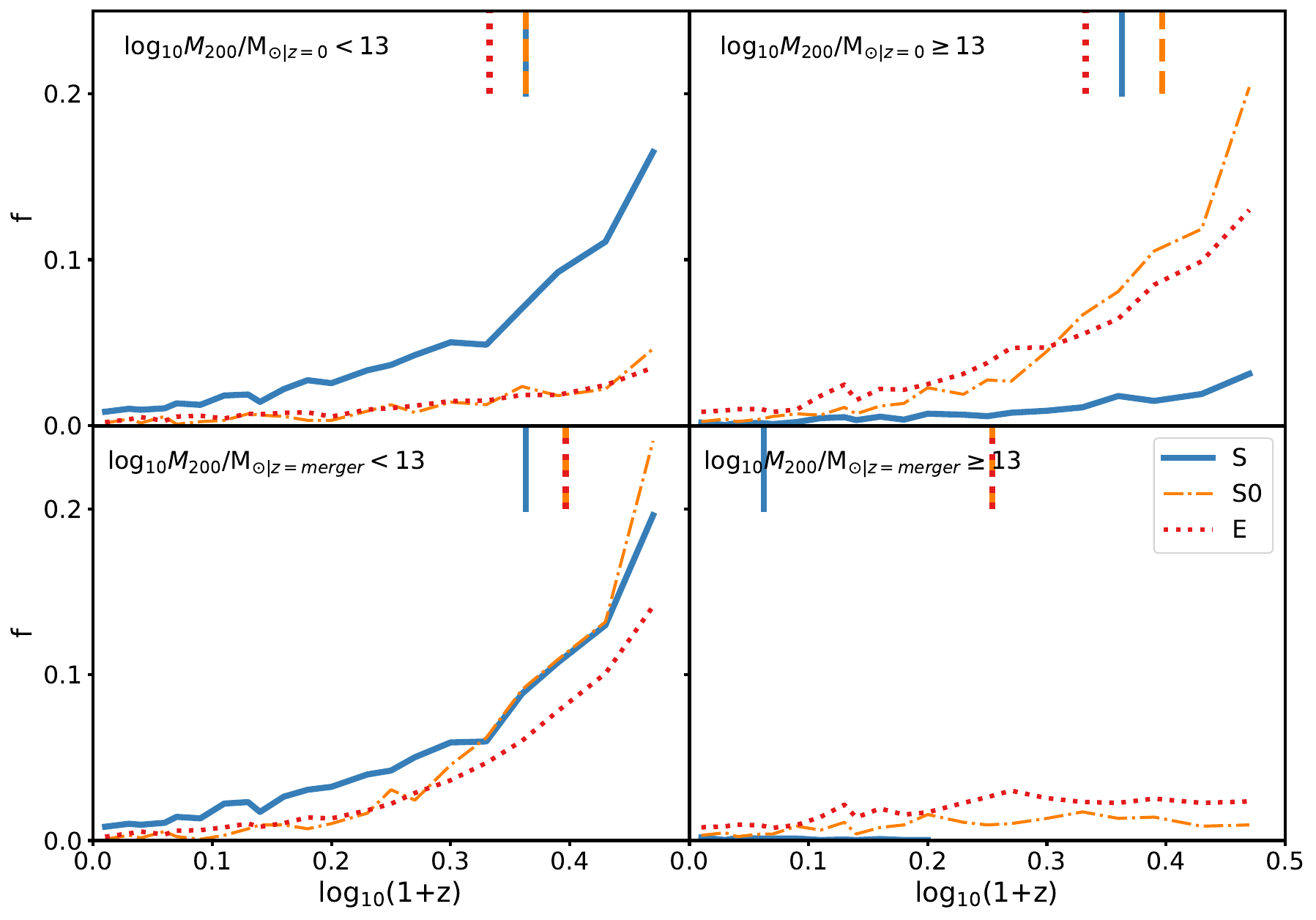}
\caption{Redshift distribution split by high-density (log$_{10}M_{200}/$M$_\odot \geq 13$; right panels) and low-density (log$_{10}M_{200}/$M$_\odot < 13$; left panels) environment. Upper and lower panels split the sample by the environment in which galaxies reside at $z=0$ (upper panels) and at $z=z_{merger}$, the time of merger (lower panels). Spirals, elliptical, and lenticular galaxies are shown in solid (blue), dotted (red), and dashed (orange) lines, respectively.}
\label{fig:z_dist_z0}
\end{figure*}

To complement this analysis, in Fig.~\ref{fig:z_dist_z0} we show the redshift distribution of said mergers, split in the same fashion, with the two left-hand panels showing galaxies in low-mass haloes (log$_{10}$M/M$_\odot < 13$), and their counterparts in more massive haloes are shown in the right-hand panels. Again, the top row classifies galaxies based on their environment at $z=0$, while the bottom row splits by their environment and the time of each merger ($z_\mathrm{merger}$). Solid (blue), dashed (orange), and dotted (red) lines show the redshift distribution of S, S0, and E galaxies, respectively.

Our analysis shows that, regardless of morphological type, most mergers occurred within the redshift range $1 < z < 2$ ($0.3 < \log_{10}(1+z) < 0.5$). Notably, the high-redshift peak in merger activity is more pronounced for lenticular galaxies than for any other morphology. This suggests that mergers involving S0s were not only predominantly minor but also took place earlier, making them less effective at significantly transforming their morphology. Spiral galaxies follow a similar trend, with merger activity occurring slightly later but still concentrated within the same redshift interval. These results are consistent with our previous findings, which indicate that S0 mergers mainly occur in low-density environments.
In contrast, elliptical galaxies exhibit a flatter redshift distribution, especially in high-density environments.

\begin{figure}
\centering
\includegraphics[width=0.5\textwidth]{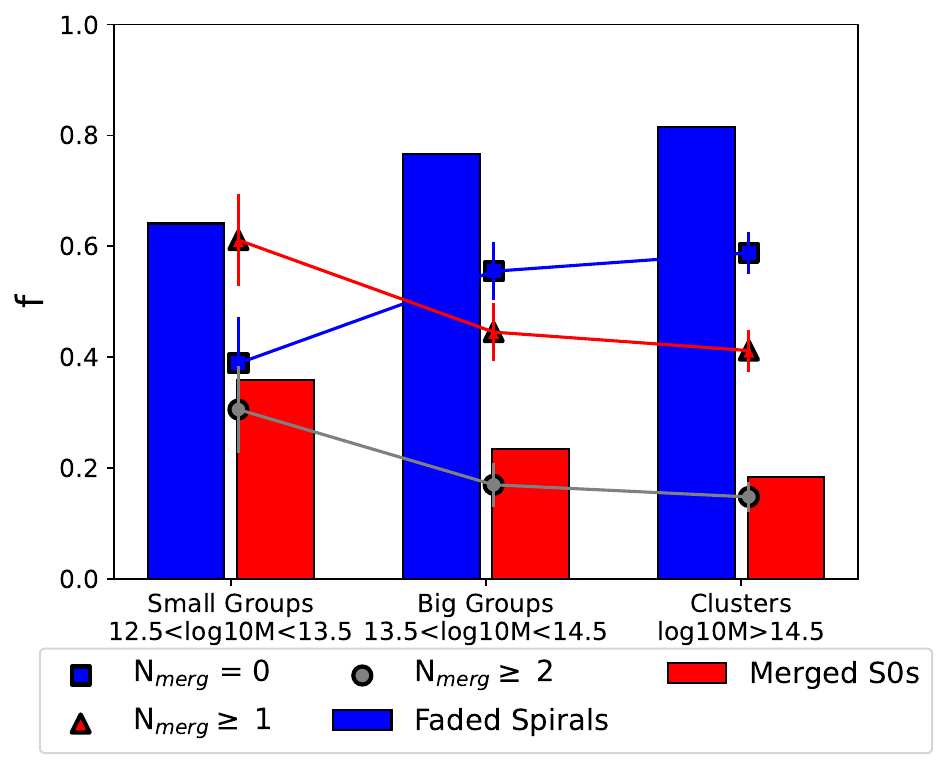}
\caption{Environmental distribution of lenticular galaxies. Blue and red bars show the distribution of faded spirals and merged-formed lenticular galaxies from the observations of \protect\citet{Coccato22}. Squares (blue), triangles(red), and circles (gray) show lenticular galaxies formed with N$_{mergers} = 0,1,\geq2$ mergers, respectively. Error bars correspond to binomial errors from the sample. As can be seen, there is a remarkable match between observations and simulations; nonetheless, the trends for merged formed S0s are clearer for galaxies in the simulation that experienced $N_{meregers}\geq2$. These results show that faded spirals dominate in dense environments and become less relevant in less dense environments.}

\label{fig:comp_lodo}
\end{figure}

\section{Discussion}

We adopt a physically motivated morphological classification combining kinematics and star-formation activity: S0s are quenched disks ($\lambda_R/\sqrt{\varepsilon} > 0.31$, $\log_{10}(\mathrm{sSFR}[yr^{-1}]) < -11$), spirals are star-forming disks (same kinematic threshold, $\log_{10}(\mathrm{sSFR}[yr^{-1}]) > -10.5$), and ellipticals are quenched spheroids ($\lambda_R/\sqrt{\varepsilon} < 0.31$, $\log_{10}(\mathrm{sSFR}[yr^{-1}]) < -11$). This kinematically informed scheme, inspired by integral-field spectroscopy studies \citep[e.g.,][]{Emsellem07, Cappellari25}, mitigates projection biases inherent in visual morphology and enables consistent comparisons across environments.
Using this scheme, Figures~\ref{fig:sat_cen-split} and~\ref{fig:M200_merg} reveal an apparent increase in S0 and elliptical fractions with host halo mass~($M_{200}$). Whether this reflects a genuine morphology--environment relation remains debated: some studies find no dependence of early-type fractions on cluster velocity dispersion \citep{Postman2005ApJ, Poggianti2009ApJ, Simard2009A&A}, while others report a clear correlation \citep{Oh2018ApJS} or a weak residual environmental signal after controlling for stellar mass \citep{vandeSande21}. Discrepancies likely reflect differences in classification methodology, and further investigation under criteria aligned with prior observational work is warranted.

S0 galaxies occupy a pivotal position between spirals and ellipticals, and whether they originate from faded spirals, merger remnants, or both remains a subject of discussion. Our results provide new constraints on this question by highlighting the critical roles of environment and merger history.
We find that over 85\% of S0s reside as satellites in groups and clusters ($\log_{10}(M_{200}/M_\odot) > 13$), supporting environmental quenching as a dominant pathway, consistent with \citet{Gort25}. Nonetheless, most merger activity in S0s occurs in low-density environments prior to cluster infall, supporting a pre-processing scenario and the delayed-then-rapid quenching framework \citep[e.g.,][]{Poci21, Gort25}. The contrast between the active merger histories of central S0s and the quiescent histories of satellites---75\% of which have experienced no significant merger since $z=2$---is consistent with the distinct bulge and disk formation histories reported by \citet{Pak21} and \citet{Barsanti21}, and with the earlier merger peak relative to spirals, which aligns with stellar age gradients in early-type subpopulations \citep{Dominguez-Sanchez20}.

The most directly comparable observational work is that of ~\citetalias{Coccato22}, who applied a K-means clustering algorithm to kinematic and photometric properties of visually classified S0s from SAMI and MaNGA. Their analysis recovers two well-separated populations: 
\begin{itemize}
    \item A population with prominent bulges, redder colours, larger sizes, higher stellar masses, and signs of past interactions, interpreted as merger-formed S0s.
    \item A second group with higher rotational support, lower metallicities, and disk-like structural features, interpreted as faded spiral galaxies.
\end{itemize}

\begin{figure}
\centering
\includegraphics[width=0.5\textwidth]{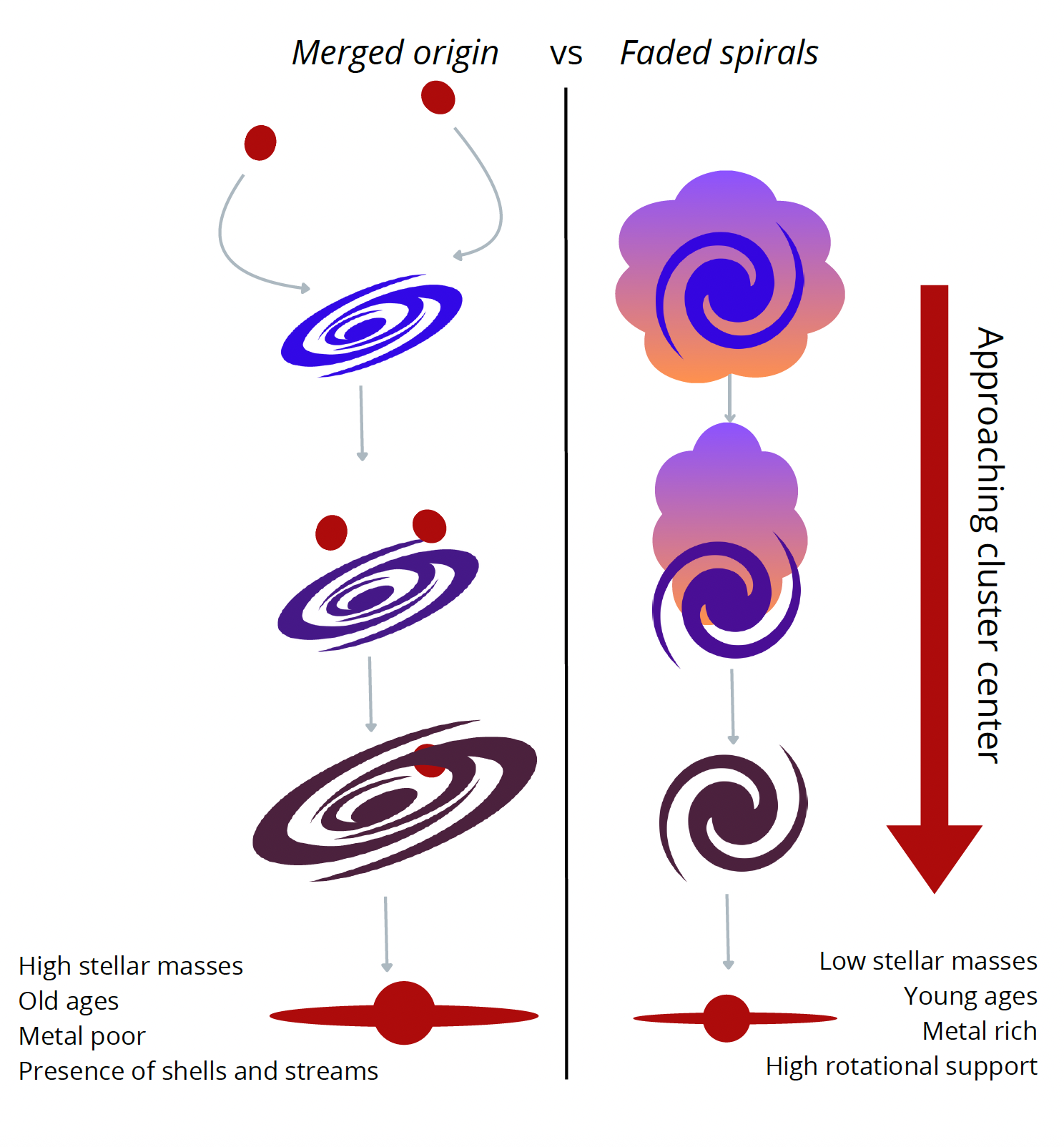}
\caption{Schematic representation of the two main formation pathways for lenticular galaxies. On one hand, S0s formed via mergers are characterised by prominent bulges, redder colours, larger sizes, and higher stellar masses. On the other, S0s formed via environmental quenching retain high rotational support, exhibit lower metallicities, and show disk-like structures.}
\label{fig:scheme_s0s}
\end{figure}

In Figure~\ref{fig:comp_lodo}, we compare the environmental distribution of our simulated S0s with \citetalias{Coccato22}. Simulated S0s with $N_{mergers}\geq 2$ closely match the observationally classified merger-formed population across all halo-mass bins, agreeing within the error bars. This supports a dual-channel scenario in which spiral fading---driven by ram-pressure stripping or strangulation---dominates in groups and clusters, while merger-driven formation is more prevalent in lower-density environments. A schematic of these pathways is shown in Figure~\ref{fig:scheme_s0s}.
Notably, we recover formation histories strikingly consistent with \citetalias{Coccato22} using only merger histories, whereas they used structural and stellar population properties. This highlights the potential of clustering algorithms to reconstruct galaxy formation pathways, and motivates future work applying such methods directly to simulated galaxies across a broader observational baseline.
Overall, our results support a picture in which S0s and spirals build most of their stellar mass through star formation and mergers in low-density environments at $1<z<2$, before migrating into denser regions. Although present-day S0s inhabit environments similar to ellipticals, their formation histories more closely resemble those of spirals--transformed via mergers in the field or small groups, then further evolved by environmental quenching in clusters. Lenticular galaxies thus represent a genuinely transitional population, structurally and environmentally, between spirals and ellipticals. Extending this analysis to a cosmological volume, such as COLIBRE \citep{Schaye25, Chaikin25}, will be essential to fully constrain the relative importance of these formation channels across all environments.

\textbf{\subsection{Caveats}}
\label{sec:caveats}

\subsubsection{Selection bias}

The Hydrangea suite targets massive clusters with $14 \leq \log_{10}(M_{200}^{z=0}/\mathrm{M}_\odot) \leq 15.4$, so the overall halo-mass distribution is not representative of a cosmological volume. As shown in Appendix~\ref{sec:TNG}, a volume-limited simulation (TNG100) yields a substantially larger fraction of S0s in low- and intermediate-density environments ($M_{200} < 10^{13}\,\mathrm{M}_\odot$), directly affecting the relative importance of the two formation channels. Our conclusions regarding the dominance of faded spirals are therefore most robust in cluster environments and should be interpreted with caution when extrapolated to field and group populations. We plan to address this limitation using cosmological-volume simulations such as COLIBRE \citep{Schaye25, Chaikin25}, and note that this study serves as a pilot for comparison with observational datasets such as SAMI and the future CHANCES survey \citep{Sifon25, Mendez-Hernandez26}.
We additionally restrict our analysis to $M_\star > 10^{10}\,\mathrm{M}_\odot$, ensuring sufficient resolution for robust kinematic and morphological measurements. This excludes the low-mass S0 population, where harassment and tidal stripping may be more prominent than ram-pressure stripping and mergers.

\subsubsection{Three-dimensional vs. projected kinematic measurements}

Our morphological classification relies on 3D stellar kinematics ($\lambda_R$, $\varepsilon$) computed from all particles within the aperture, whereas observers derive these quantities from projected 2D light maps. Previous work has shown that 3D mass-weighted $\lambda_R$ yields consistent classifications with its projected analogue \citep{Naab14, Lagos17, Pallero25}, though projection effects can scatter individual galaxies across the $\lambda_R$ boundary. Similarly, our non-iterative inertia-tensor ellipticities may be systematically underestimated in flattened systems, though this has little impact on the statistical class distributions \citep{Pallero25}.
Our fixed thresholds in $\lambda_R/\sqrt{\varepsilon}$ and sSFR are validated in previous work \citep{Emsellem11, Naab14, Lagos17, Pallero25} and robust within $\pm0.25$ in sSFR \citep{Pallero22}, but may not perfectly map onto observational classification schemes, which are predominantly visual or photometric. This is relevant for the comparison with \citetalias{Coccato22}, whose S0 sample was visually selected; differences in strategy may introduce a systematic offset in the relative fractions of faded spirals and merger-formed S0s, though the overall agreement in Figure~\ref{fig:comp_lodo} suggests this effect is small.

\subsubsection{Merger definition and comparison with observations}

We classify S0s as merger-formed based on whether they experienced at least one merger with $f_i > 1:10$ since $z=2$. Interactions below this threshold can still induce structural changes, so the boundary between populations is not perfectly sharp. Additionally, \textsc{SPIDERWEB} may miss short-lived or poorly resolved interactions, meaning our estimate of $\sim\,75\%$ faded spirals among satellite S0s should be treated as an upper limit.
Our comparison with \citetalias{Coccato22} is a qualitative consistency check: the observational sample is small, and the classification criteria---based on stellar populations, kinematics, and photometry---are not directly equivalent to our merger-history approach. Future work applying clustering algorithms to simulated galaxy properties will enable a more rigorous test.

\section{Summary and Conclusions}

\label{sec:summary}

In this work, we explored the formation pathways of lenticular (S0) galaxies located within 10$r_{200}$ of galaxy clusters, using high-resolution hydrodynamical simulations. Our primary goal was to determine whether S0 galaxies predominantly originate from faded spiral galaxies or through merger-driven processes, and how these pathways vary with the environment. To do so, we analysed a sample of galaxies from cosmological simulations, studying their merger histories, morphological transformations, and environmental dependencies. Our main findings are as follows:   
\begin{enumerate}
      \item Lenticular galaxies reside preferentially in high-density environments (log$_{10}M_{200}/$M$_\odot$ > 13) such as groups and clusters, being predominantly ($>85\%$) satellites of these structures. On the other hand, a small fraction ($\sim 10\%$) of lenticular galaxies are identified as central galaxies from less massive haloes. 
      \item When looking at the merging history of lenticular galaxies, the satellite population is characterised by a very quiescent merging history, with $\sim 75\%$ of them having no significant mergers at $z<2$ and with $<10\%$ having more than two significant mergers at $z<2$. Lenticular galaxies at the centres of their haloes show a mixed merging history, ranging from none to several. Nevertheless, we remind the reader that only a small percentage ($<15\%$) of lenticular galaxies reside at the centre of low-density environments.
      \item For galaxies that did undergo mergers, we examined the distribution of mass ratios and found that spiral galaxies tend to experience the most minor mergers, followed by S0s. When split by environment, S0s show a clear trend of merging in low-density environments, even if they currently reside in dense clusters. This indicates that their transformations often occurred before infall, supporting the preprocessing scenario.
      \item Most mergers occurred between $1<z<2$, with S0s showing an earlier merger epoch than spirals, particularly in low-density environments. Elliptical galaxies, on the other hand, exhibit a flatter distribution, especially in clusters, pointing toward ongoing assembly at lower redshifts.
      \item When comparing our simulated populations with the observational study of \citetalias{Coccato22}, who used a K-means clustering algorithm on visually classified lenticulars from the SAMI and MaNGA surveys, our simulation-based classifications show remarkable agreement with their results, particularly when selecting S0s that experienced two or more mergers at $z<2$. Both simulations and observations agree that faded spirals are the dominant formation channel, especially in high-density environments, while merger-driven S0s become more relevant in lower-density environments.
   \end{enumerate}

Our findings provide strong evidence that spiral fading is the primary evolutionary route for lenticular galaxies, especially in cluster environments, while mergers contribute significantly in groups and the field. Our results underscore the importance of environment and cosmic time in shaping galaxy morphology and demonstrate the power of combining simulations and observations to uncover the physical processes behind galaxy evolution.
In future work, we aim to integrate machine learning clustering techniques into simulation-based galaxy selection, enabling more detailed and direct comparisons with spectroscopic and photometric surveys such as SAMI \citep{SAMI}, MaNGA \citep{Manga}, and S-PLUS \citep{MendesdeOliveira19}.

\begin{acknowledgements}
       DP acknowledges support from the ESO Early-Career Scientific Visitor Programme. YLJ and DP acknowledge support from the Agencia Nacional de Investigación y Desarrollo (ANID) through Basal project FB210003, FONDECYT Regular projects 1241426 and 123044, and  Millennium Science Initiative Program NCN2024\_112. 
      AFM gratefully acknowledges the sponsorship provided by the European Southern Observatory through a research fellowship.
      FAG acknowledges support from the ANID BASAL project FB210003, from the ANID FONDECYT Regular grants 1251493, and the HORIZON-MSCA-2021-SE-01 Research and Innovation Programme under the Marie Sklodowska-Curie grant agreement number 101086388.
      E.J.J. acknowledges support by the ANID BASAL project  FB210003. CL-D acknowledges financial support from the ESO Comit\'e Mixto 2022. AC acknowledges the Fundação de Amparo à Pesquisa do Estado do Rio de Janeiro (FAPERJ) grant E26/202.607/2022 e 210.371/2022(270993) and the 'Bolsa de Produtividade do CNPQ nível 2'. MEDR acknowledges support from {\it Agencia Nacional de Promoci\'on de la Investigaci\'on, el Desarrollo Tecnol\'ogico y la Innovaci\'on} (Agencia I+D+i, PICT-2021-GRF-TI-00290, Argentina).
\end{acknowledgements}

%
%

\bibliographystyle{aa} 
\bibliography{quench_final} 

\begin{appendix}
\section{Merger History and Environmental Trends in TNG100}
\label{sec:TNG}

\begin{figure}
\centering
\includegraphics[width=0.5\textwidth]{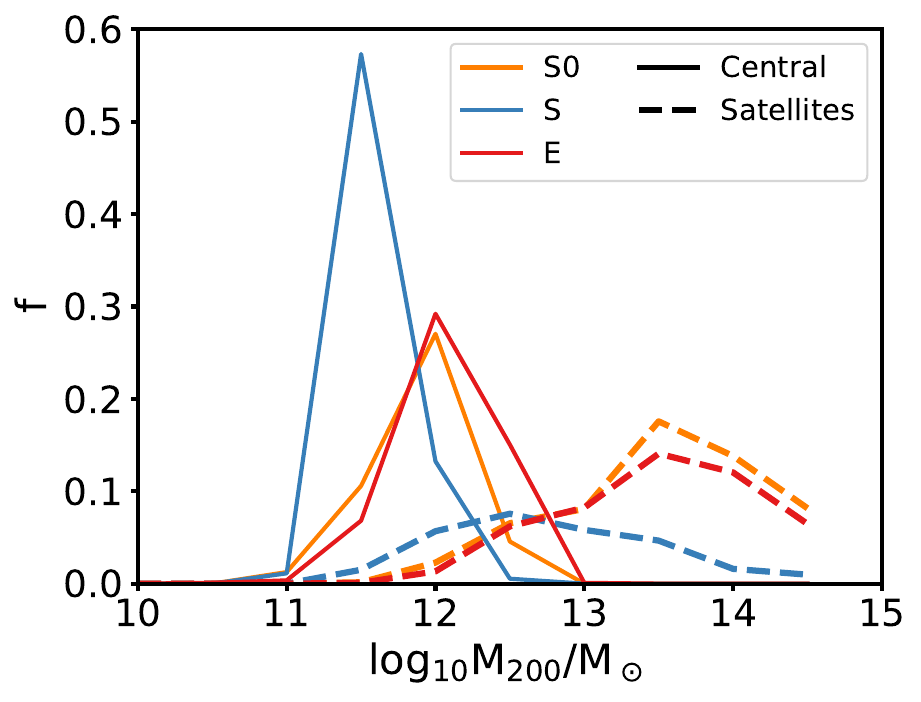}
\caption{Central–satellite split for galaxies in the TNG100 simulation. Red, blue, and orange lines represent elliptical, spiral, and lenticular populations, respectively. Solid and dashed lines indicate central and satellite galaxies. Spiral galaxies mainly reside as centrals in low-mass haloes, while lenticulars and ellipticals are more evenly distributed, with a preference for being satellites in haloes with $\log_{10}(M_{\star}/{\rm M_\odot}) \ge 13$. The overall distributions of ellipticals and lenticulars are remarkably similar.}

\label{fig:satcen_TNG}
\end{figure}

\begin{figure}
\centering
\includegraphics[width=0.5\textwidth]{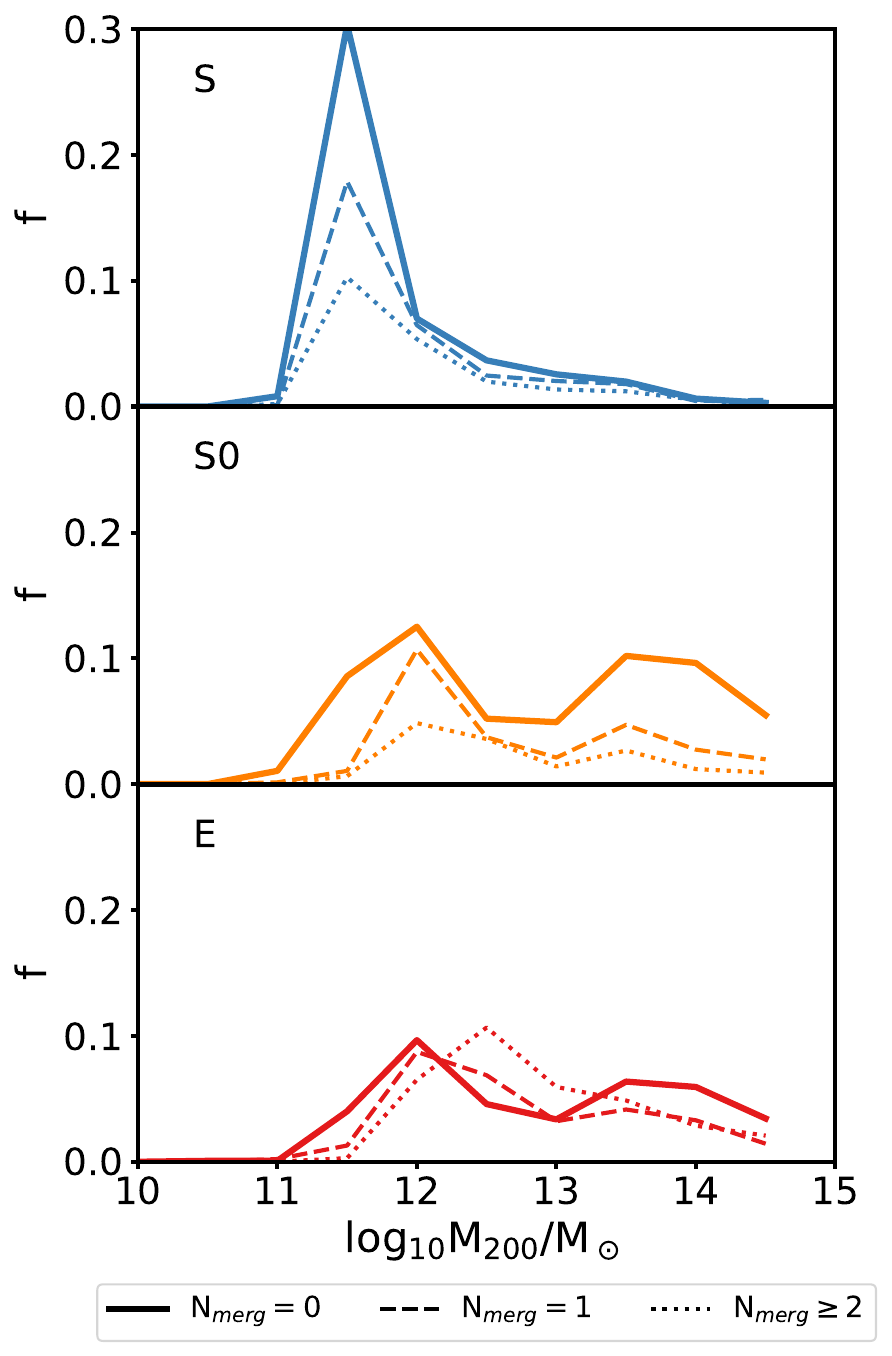}
\caption{Halo-mass distributions for galaxies of different morphologies, split according to the number of significant mergers experienced since $z=2$ in the full IllustrisTNG-100 simulation. Galaxies are classified as spirals (top, blue), lenticulars (middle, orange), and ellipticals (bottom, red). Solid, dashed, and dotted lines correspond to galaxies that experienced no mergers, one merger, and two or more mergers, respectively.
Spiral galaxies display similar halo-mass distributions regardless of merger history. Lenticular galaxies without mergers are preferentially found in massive haloes, whereas those that experienced mergers tend to inhabit lower-mass systems. Ellipticals exhibit broad halo-mass distributions, consistent with diverse formation pathways.}

\label{fig:M200_TNG}
\end{figure}

Figures \ref{fig:satcen_TNG} and \ref{fig:M200_TNG} illustrate the environmental dependence of galaxy morphology and merger history within the TNG100 simulation. Figure \ref{fig:satcen_TNG} shows the distribution of central and satellite galaxies as a function of halo mass for different morphological types. Spiral galaxies predominantly reside as centrals in low-mass haloes, whereas lenticular and elliptical galaxies display more balanced central--satellite fractions, with satellites preferentially occupying haloes with $\log_{10}(M_{200}/{\rm M_\odot}) \gtrsim 13$. The similarity between the lenticular and elliptical distributions suggests that dense environments strongly influence both populations.

Figure \ref{fig:M200_TNG} further explores the connection between morphology, halo mass, and merger history. For each morphological type, we show the halo-mass distribution, split by the number of significant mergers experienced since $z=2$, for the full sample (left panels). Spiral galaxies exhibit similar halo-mass distributions regardless of merger history. In contrast, E/S0 galaxies show two prominent peaks at low ($M_{200}/M_{\odot} \sim 10^{12}$) and intermediate ($M_{200}/M_{\odot} \sim 10^{13.5}$) halo masses, with the distribution declining towards more massive systems. These galaxies also exhibit a broad range of merger histories across all environments, highlighting the diverse evolutionary pathways of early-type galaxies.

\end{appendix}

\end{document}